\documentclass[conference]{IEEEtran}
\IEEEoverridecommandlockouts

\usepackage{cite}
\usepackage{amsmath,amssymb,amsfonts}
\usepackage[ruled,linesnumbered]{algorithm2e}
\usepackage{graphicx}
\usepackage{makecell}
\usepackage{multicol}
\usepackage{multirow}
\usepackage[table]{xcolor}  
\usepackage{colortbl}
\usepackage{multirow}
\usepackage{textcomp}
\usepackage{xcolor}
\def\BibTeX{{\rm B\kern-.05em{\sc i\kern-.025em b}\kern-.08em
    T\kern-.1667em\lower.7ex\hbox{E}\kern-.125emX}}

\newtheorem{definition}{Definition}

\newtheorem{example}{Example}

\begin{document}

\title{Reducing ORBGRAND Latency via Partial Gaussian Elimination\\

\thanks{This work was supported in part by the National Natural Science Foundation of China under Grant 62231022.}
}

\author{%
\IEEEauthorblockN{Li Wan, Wenyi Zhang}
\IEEEauthorblockA{Department of Electronic Engineering and Information Science\\
                   University of Science and Technology of China\\
                   Email: wan\_li@mail.ustc.edu.cn, wenyizha@ustc.edu.cn}
}

\maketitle

\begin{abstract}
    Guessing Random Additive Noise Decoding (GRAND) is a universal framework for decoding all block codes by testing candidate error patterns (EPs). Ordered Reliability Bits GRAND (ORBGRAND) facilitates parallel implementation of GRAND by exploiting log-likelihood ratio (LLR) rankings but still suffers from high tail latency under unfavorable channel conditions, limiting its use in real-time systems.

    We propose an elimination-aided ORBGRAND scheme that reduces decoding latency by integrating the Rank of the Most Reliable Erroneous (RMRE) bit with a partial Gaussian-elimination (GE) filtering mechanism. The scheme groups and jointly verifies EPs that share the same RMRE, and once a valid EP is identified, the ORBGRAND search is resumed. By leveraging prior GE steps to filter out unnecessary guesses, this approach significantly reduces the number of EPs to be tested, thereby lowering both average and worst-case latency while maintaining error-correction performance.

    Simulation results show that compared to the original ORBGRAND, the elimination-aided ORBGRAND filters out more than 50\% of EPs and correspondingly reduce overall computational complexity, all with no loss in block error rate. This demonstrates that this approach is suitable for ultra-reliable low-latency communication scenarios.
\end{abstract}


\section{Introduction}

In recent years, a variety of decoding methods collectively referred to as Guessing Random Additive Noise Decoding (GRAND) have been proposed. These methods are often termed universal
in that they can be applied to all block codes (e.g. BCH, LDPC, and Polar codes). The core idea is to sequentially subtract candidate error patterns (EPs) from the noisy channel output to determine whether a valid codeword is recovered. GRAND is particularly suitable for short, high-rate codes \cite{duffy2019capacity}, making it an attractive candidate for ultra-reliable low-latency communication (URLLC) \cite{yue2023efficient}.

Among GRAND variants, Ordered Reliability Bits GRAND (ORBGRAND) generates queries based only on the ranking of log-likelihood ratio (LLR) magnitudes \cite{duffy2022ordered, liu2022orbgrand}, facilitating parallelism and hardware-friendly implementation \cite{abbas2022high, ji2024efficient}, \cite{wan2025parallelism}. As a result, ORBGRAND represents one of the promising solutions towards realizing low-latency communication. However, ORBGRAND decoders remain vulnerable to latency instability: although the average decoding time is often low, the worst-case number of guesses can increase dramatically under unfavorable channel conditions, resulting in tail latency that may be unacceptable in real-time systems.

Previous studies on accelerating ORBGRAND can be roughly divided into two categories. The first focuses on designing specific pre-generated EP sets to achieve better decoding performance, which typically reduces both the average number of guesses and overall complexity \cite{duffy2022ordered, liu2022orbgrand, wan2024approaching}.
The second aims at designing faster verification schemes to reduce the decoding delay: some approaches leverage the structure of the parity-check matrix to eliminate certain EPs early \cite{rowshan2022constrained} or accelerate syndrome-based searches \cite{hadavian2023ordered}, while others generate EPs in real time by partially evaluating check equations \cite{wang2024partially}. 

We introduce a new perspective based on the Rank of the Most Reliable Erroneous (RMRE) — the index of the most reliable bit flipped in an EP. EPs with smaller RMRE indices are statistically more likely to occur, dominating the early stages of decoding. Building on this observation, we incorporate partial Gaussian elimination (GE) into the ORBGRAND decoding process to jointly verify multiple EPs sharing the same RMRE, thereby accelerating decoding.  

To implement the elimination-aided ORBGRAND for a code with information length $K$ and codeword length $N$, the decoder reorders the parity-check matrix $H^{(N-K)\times N}$ according to bit reliability and then applies partial GE column by column, starting from the least reliable bit. A similar elimination procedure is often employed in ordered-statistics decoding (OSD) \cite{fossorier1995soft}, but in our method, the elimination is optimized to reduce computational complexity. Specifically:
\begin{itemize}
    \item At each column $k$, the decoder checks whether a codeword can be obtained by flipping bits within the first $k$ positions. This allows all EPs with RMRE $=k$ to be verified jointly, thereby filtering out redundant guesses.
    \item The elimination targets only the first $k$ columns, ignoring irrelevant remaining columns to reduce complexity.
\end{itemize}

These two operations reduce the complexity of GE from $\mathcal{O}(N(N-K)^2)$ to $\mathcal{O}((N-K)M^2)$, where $M$ (typically 20-30\% of $N-K$) represents the number of columns required for partial GE, significantly reducing computational overhead.
Once a valid EP is identified within the GE module, the decoder returns to the ORBGRAND sequence and proceeds with decoding, safely skipping all EPs with RMRE less than or equal to the determined EP. This strategy significantly reduces the total number of guesses and accelerates the decoding.

Simulation and complexity analyses demonstrate that the elimination-aided ORBGRAND  accelerates decoding without compromising error-correction performance. The method is hardware-friendly, relying exclusively on bitwise XOR operations, and the embedded partial GE enables seamless integration with OSD for rare, hard-to-decode cases. Together, these features provide a scalable and efficient framework for GRAND-based architectures.

The remainder of the paper is organized as follows: Section~\ref{sec:system_model} introduces the transmission model and the ORBGRAND, Section~\ref{sec:results} details the elimination-aided ORBGRAND and performs a complexity analysis, Section~\ref{sec:performance} presents simulation results validating the effectiveness of the proposed approach. Finally Section~\ref{sec:conclusion} concludes this paper.

\section{Preliminaries}\label{sec:system_model}

\subsection{Transmission Model}\label{2.1}

We consider a block coding system where an information vector $\underline{U}\in\mathbb{F}_{2}^{K}$ is encoded into a codeword $\underline{W}\in\mathcal{C}\subseteq\mathbb{F}_{2}^{N}$ with code rate $R=K/N$. The codeword is mapped to a transmitted signal vector $\underline{X}\in\{-1,+1\}^{N}$ by antipodal modulation, where $X_i=+1$ if $W_i=0$ and $X_i=-1$ if $W_i=1$. The resulting AWGN channel output vector satisfies $\underline{Y}\mid\underline{X} \sim \mathcal{N}(\underline{X}, \sigma^2 \mathbf{I}_{N \times N})$. Given $\underline{Y}$, the LLRs are given by
\begin{equation}\label{eq:LLR}
    L_{i} = \log\frac{p(Y_i \mid W_{i} = 0)}{p(Y_i \mid W_{i} = 1)} = \frac{2}{\sigma^2} Y_i,\ i = 1, \ldots, N,
\end{equation}
and we use $\ell_i$ to represent the realization of $L_i$.

We introduce the hard decision function $\theta(\cdot)$ as $\theta(y) = 0$ if $y \geq 0$ and $1$ otherwise. For $\underline{Y}$, denote the vector of $[\theta(Y_1), \ldots, \theta(Y_n)] \in \mathbb{F}_2^N$ by $\theta(\underline{Y})$ for simplicity.

\subsection{ORBGRAND}\label{subsec:GRAND}

GRAND consists of two components: an EP generator and a codeword checker. Given a maximum query budget $T$, the decoder produces a sequence of EPs $\{\underline{e}(1), \dots, \underline{e}(T)\}$ with $\underline{e}(t)\in\mathbb{F}_{2}^{N}$, and sequentially test whether $\theta(\underline{y}) \oplus \underline{e}(t)$ is a valid codeword. For linear block codes with check matrix $H^{(N-K)\times N} = \left[ \underline{h}_1,...,\underline{h}_{N}\right]$, this test can be expressed as
\begin{equation}\label{eq:check}
H \cdot \left(\theta(\underline{y}) \oplus \underline{e}(t)\right) = \underline{0} 
\quad \Leftrightarrow \quad
H \cdot \underline{e}(t) = \underline{s} \ , 
\end{equation}
where we define the syndrome $\underline{s} \triangleq H \cdot \theta(\underline{y})$. If the resulting $\theta(\underline{y}) \oplus \underline{e}(t)$ corresponds to a valid codeword, decoding terminates with this output, and we call such an EP valid; otherwise the process continues until either a codeword is found or the budget $T$ is exhausted, in which case decoding is declared unsuccessful.

ORBGRAND is a variant GRAND algorithm that employs an efficient EP generation strategy. Specifically, it precomputes and locally stores a set of EPs, denoted as $\mathcal{E}_{\text{ORB}} = \{\tilde{\underline{e}}(1), \dots, \tilde{\underline{e}}(T)\}$\footnote{We use $\underline{e}$ for the actual guessed EP, while $\tilde{\underline{e}}$ refers to the locally pre-stored EP in ORBGRAND. The two are related through the transformation in \eqref{eq:permute}.}. For each received LLR vector $\underline{\ell}$, the elements are sorted to obtain a ranking vector $\underline{r}$ such that
\begin{equation}\label{eq:ranking}
    |\ell_{r_1}| \le |\ell_{r_2}| \le \dots \le |\ell_{r_N}|.
\end{equation}
Each pre-stored EP $\tilde{\underline{e}}(t)$ is then permuted according to $\underline{r}$ prior to testing, producing
\begin{equation}\label{eq:permute}
    \underline{e}(t) = \mathcal{P} \cdot \tilde{\underline{e}}(t),
\end{equation}
where the permutation matrix $\mathcal{P} \in \{0,1\}^{N\times N}$ is defined as $\mathcal{P}_{i,j} = \mathbf{1}(i = r_j)$. Consequently, the check in \eqref{eq:check} can be rewritten as
\begin{equation}\label{eq:check_2}
H \cdot (\mathcal{P} \cdot \tilde{\underline{e}}(t)) = \underline{s}.
\end{equation}
By grouping $H$ and $\mathcal{P}$, we define $H_{\pi} \triangleq H \cdot \mathcal{P}$, yielding
\begin{equation}\label{eq:permutation_check}
    H_{\pi} = \left[ \underline{h}_{r_1}, \dots, \underline{h}_{r_N}\right], \quad \text{Eq. }\eqref{eq:check_2} \ \Leftrightarrow \ H_{\pi} \cdot \tilde{\underline{e}}(t) = \underline{s}.
\end{equation}

For clarity, the overall decoding flow of ORBGRAND is summarized in \textbf{Algorithm~\ref{alg:ORBGRAND}}.
\begin{algorithm}[htbp]
    \caption{ORBGRAND \cite{duffy2022ordered}}\label{alg:ORBGRAND}
    \KwIn{$\underline{y}$, $T$, $H$, pre-stored $\mathcal{E}_{\text{ORB}}$.}
    \KwOut{Decoding result $\hat{\underline{w}}$.}
    $\mathcal{P} \gets \text{Sorting } \underline{y}$, initialize $t = 0$, $flag = 0$\;
    $s \gets H \cdot \theta(\underline{y}),\ H_{\pi} = H \cdot \mathcal{P}$\;
    \While{$flag = 0$ \textbf{and} $t < T$}{
        $t = t + 1$\;
        $\tilde{\underline{e}}(t) \gets \mathcal{E}_{\text{ORB}}[t]$\;
        \textit{flag} $\gets$ $\mathbf{1}(H_\pi \cdot \tilde{\underline{e}}(t) = \underline{s})$ \tcp*{\textit{flag} = 1 if successful}
    }
    \Return $\hat{\underline{w}} = 
    \begin{cases} 
        \theta(\underline{y})\oplus (\mathcal{P}\cdot \tilde{\underline{e}}(t)) & \text{if \textit{flag} = 1}, \\
        \emptyset & \text{if \textit{flag} = 0}.
    \end{cases}$
\end{algorithm}

\subsection{Rank of the Most Reliable Erroneous}\label{subsec:rmre}

In this work, we focus on the error bit with the highest reliability within a codeword.  
To formalize this concept, we first define the Rank of the Most Reliable Erroneous (RMRE) for an EP.

\begin{definition}
    Let $\underline{e} \in \mathbb{F}_2^n$ denote an EP, and let $\underline{r}$ be the reliability ranking vector as in \eqref{eq:ranking}.  
    The RMRE of $\underline{e}$ is defined as
    \begin{equation}\label{eq:rmre}
        \text{RMRE}(\underline{e}) = \max \left\{ i \mid e_{r_i} = 1 \right\}.
    \end{equation}
\end{definition}

In particular, the RMRE of the unique EP that can lead to successful decoding can be expressed as
\begin{align}
    \text{RMRE}(\theta(\underline{y}) \oplus \underline{w}) 
    &= \max \left\{ i \mid \theta(y_{r_i}) \oplus w_{r_i} = 1 \right\} \notag \\
    &= \max \left\{ i \mid \theta(y_{r_i}) \neq  w_{r_i} \right\}. \label{eq:rmre_2}
\end{align}
And for ORBGRAND, since each tested EP is written as $\underline{e} = \mathcal{P} \cdot \tilde{\underline{e}}$, with the definition of $\mathcal{P}$, we have
\begin{equation*}
    \text{RMRE}(\mathcal{P}\cdot \tilde{\underline{e}}) = \max \left\{ i \mid \tilde{e}_i = 1 \right\}.
\end{equation*}

\section{Main Results}\label{sec:results}

This section presents the main results of this paper. Section \ref{subsec:algorithm} provides a detailed description of the algorithm flow, Section \ref{subsec:complexity} provides a complexity analysis and proposes further optimizations to the algorithm. Section \ref{subsec:discuss} further provides some analysis of RMRE for helping understand the mechanism underlying the proposed algorithm.

\subsection{Elimination-aided ORBGRAND Algorithm}\label{subsec:algorithm}

The proposed elimination-aided ORBGRAND algorithm aims to find a solution of $H_{\pi}\,\tilde{\underline{e}} = \underline{s}$ by performing partial GE, thus enabling more efficient execution of ORBGRAND.
The overall process is presented below. Due to iterative matrix transformations involved, we use superscripts (e.g., $H_{\pi}^{(k)}$) to denote the results after the $k$-th round.

\paragraph{\textbf{Step 1. Initialization}}  
Form the augmented matrix $[H_{\pi}^{(0)} \mid \underline{s}^{(0)}] \in \mathbb{F}^{(N-K) \times (N+1)}$.  
Initialize the candidate threshold as $t^{*} = T$, representing the maximum index of EPs.
Set the column index $n=1$ and the pivot index $k=1$.

\paragraph{\textbf{Step 2. Column-wise elimination}}  
For column $n$, search for a pivot among $k,\dots,N-K$.  

\textbf{Case I:} If a nonzero entry $h_{pn}^{(k-1)}$ ($k \leq p \leq N-K$) is found, swap rows $p$ and $k$ to establish a pivot at $(k,n)$, and perform elimination for all other rows:
\begin{gather*}
    \forall\ i \neq k \text{ and } \forall\ j: \quad m_{in}^{(k-1)} = \frac{h_{in}^{(k-1)}}{h_{kn}^{(k-1)}},  \\[2pt]
    h_{ij}^{(k)} \leftarrow h_{ij}^{(k-1)} - m_{in}h_{kj}^{(k-1)}, s_i^{(k)} \gets s_i^{(k-1)} - m_{in}^{(k-1)}s_k^{(k-1)}.
\end{gather*}

Since $H_{\pi}^{(k-1)}$ is binary and $h_{kn}^{(k-1)}=1$, the multiplier simplifies to $m_{in}^{(k-1)}=h_{in}^{(k-1)}$, yielding
\begin{equation}\label{eq:elimination}
    h_{ij}^{(k)} \leftarrow h_{ij}^{(k-1)} \oplus h_{in}^{(k-1)}h_{kj}^{(k-1)}, \ s_i^{(k)} \leftarrow s_i^{(k-1)} \oplus h_{in}^{(k-1)}s_k^{(k-1)}.
\end{equation}
After elimination, update $k \gets k+1$, and proceed to \textbf{\textit{Step~3}} for consistency checking.

\textbf{Case II:} If no pivot is found, update $n \gets n+1$, and continue to \textbf{\textit{Step~2}}.

\paragraph{\textbf{Step 3. Reduced-system verification}}
Whenever a pivot is found, check whether there is a solution in the first $n$ columns. This is equivalent to verifying
\begin{equation}\label{eq:consistency_check}
    \mathrm{rank}\!\left([\underline{h}_1^{(k)},\dots,\underline{h}_n^{(k)}]\right)
    =
    \mathrm{rank}\!\left([\underline{h}_1^{(k)},\dots,\underline{h}_n^{(k)},\underline{s}^{(k)}]\right),
\end{equation}
which indicates that there exists an $\tilde{\underline{e}}$ that flips only the first $n$ bits and satisfies $H_\pi^{(0)} \cdot \tilde{\underline{e}} = \underline{s}^{(0)}$.

If \eqref{eq:consistency_check} holds, solve the reduced system to obtain all valid EPs with $\text{RMRE}=n$, denoted as $\mathcal{E}_{n}$, and proceed to \textbf{\textit{Step~4}}; otherwise, update $n \gets n + 1$ and return to \textbf{\textit{Step~2}}.

\paragraph{\textbf{Step 4. Final solution search}}
Returning to ORBGRAND algorithm, we have established that:
\begin{itemize}
    \item Flipping only the first $n-1$ bits does not yield a valid codeword.
    \item The valid EPs with $\text{RMRE}=n$ have been identified, and their positions in $\mathcal{E}_{\text{ORB}}$ can be retrieved via hash lookup. If $\mathcal{E}_n \cap \mathcal{E}_{\text{ORB}} \neq \emptyset$, a solution is obtained as
    \vspace{-0.1em}
    \begin{equation}
        t^{*} = \min\{ t \mid \tilde{\underline{e}}(t) \in \mathcal{E}_n \cap \mathcal{E}_{\text{ORB}} \}.
    \end{equation}
    Otherwise, $t^{*} = T$ is still maintained.
\end{itemize}

Therefore, only EPs with $\text{RMRE} > n$ are further examined by checking
\begin{equation}
    H_{\pi}^{(0)}  \cdot \tilde{\underline{e}}(t) = \underline{s}^{(0)}.
\end{equation}
Since the EPs are pre-generated, their RMRE values can be stored in advance, avoiding recomputation during decoding.

In summary, the proposed algorithm performs GE in a column-wise incremental manner and checks consistency at each stage. In the final stage, it filters out infeasible candidate EPs before they are passed to the ORBGRAND search.
The complete pseudocode is presented in \textbf{Algorithm~\ref{alg:filter_GRAND}}. 
\begin{algorithm}[h]
\caption{Elimination-aided ORBGRAND}
\label{alg:filter_GRAND}
\KwIn{$\underline{y}$, $T$, $H$, pre-stored $\mathcal{E}_{\text{ORB}}$.}
\KwOut{Decoding result $\hat{\underline{w}}$.}

Form $[H_{\pi} \mid \underline{s}]$ with \eqref{eq:check}-\eqref{eq:permutation_check}, initialize $k \gets 1$, $t^{*} \gets T$\;

\For{$n \gets 1$ \KwTo $N$}{
    Find a row $p \in \{k,\dots,N\!-\!K\}$ such that $h_{pn} = 1$\;
    \If{such $p$ exists}{
        Swap rows $p$ and $k$ (pivot at $(k,n)$)\;
        \For{$i \neq k$ and $\forall\ j$}{
            $h_{ij}^{(k)} \leftarrow h_{ij}^{(k-1)} \oplus h_{in}^{(k-1)}h_{kj}^{(k-1)}$\;
            $s_i^{(k)} \leftarrow s_i^{(k-1)} \oplus h_{in}^{(k-1)}s_k^{(k-1)}$\;
        }
        \If{\eqref{eq:consistency_check} holds}{
            $\mathcal{E}_n \gets$ valid EPs using first $n$ columns\;
            \If{$\mathcal{E}_n \cap \mathcal{E}_{\text{ORB}} \neq \emptyset$}{
                $t^{*} \gets \min\{ t \mid \tilde{\underline{e}}(t) \in \mathcal{E}_n \cap \mathcal{E}_{\text{ORB}} \}$\;
            }
            \textbf{break}\;
        }
        $k \gets k+1$\;
    }
}

\For{$t \gets 1$ \KwTo $t^{*}\!-\!1$}{
    \If{$\text{RMRE}(\tilde{\underline{e}}(t)) > n$ \textbf{and} $H_\pi \cdot \tilde{\underline{e}}(t) = \underline{s}$}{
        \Return{$\theta(\underline{y})\oplus (\mathcal{P} \cdot \tilde{\underline{e}}(t))$}\;
    }
}

\Return{$\theta(\underline{y})\oplus (\mathcal{P} \cdot \tilde{\underline{e}}(t^{*}))$}\;
\end{algorithm}

\begin{example}\label{ex:1}
For clarity in demonstrating the proposed algorithm, we use the $(7,4)$ Hamming code as an example. Assume that after receiving $\underline{\ell}$, we calculate the syndrome $\underline{s}=(0,0,1)^{\mathsf{T}}$ and the sorted check matrix as follows:
\begin{equation*} 
H_{\pi} = 
    \begin{bmatrix} 
    1 & 1 & 1 & 0 & 0 & 0 & 0 \\ 
    0 & 1 & 1 & 1 & 0 & 1 & 0 \\ 
    1 & 0 & 1 & 1 & 1 & 0 & 1 
    \end{bmatrix}. 
\end{equation*} 
\end{example}
\begin{figure*}
    \centering
    \includegraphics[width=0.98\linewidth]{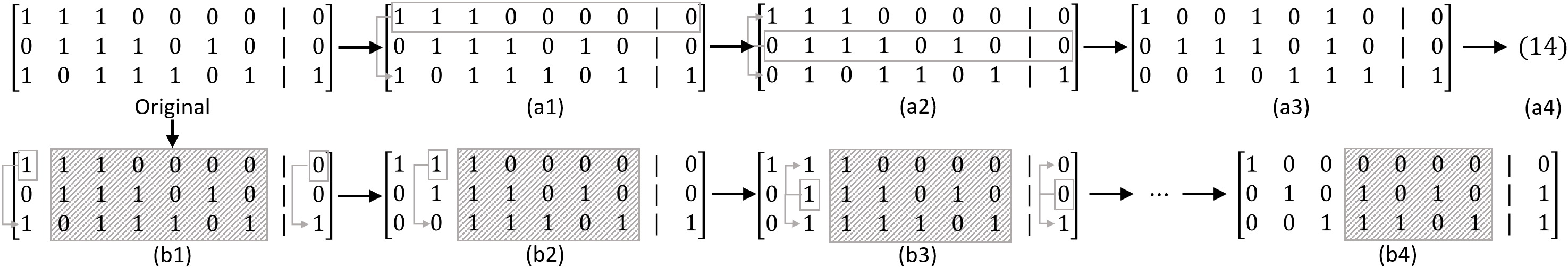}
    \caption{Comparison of the full and simplified GE schemes.  
(a1) - (a4): full elimination where each step updates all remaining columns.  
(b1) - (b4): proposed reduced elimination — (b1) processes only the first column together with $\underline{s}$.
(b2) - (b3) apply the prior transformation $(I+Q^{(1)})$ and then eliminate the second column,  
and (b4) shows the resulting reduced GE, which operates only on the first three columns and $\underline{s}$, leaves all subsequent columns untouched.}
    \label{fig:GE}
\end{figure*}
Then the corresponding augmented matrix is
\begin{equation}\label{eq:matrix_0}
    [H_{\pi}^{(0)} \mid \underline{s}^{(0)}] = 
    \begin{bmatrix} 
    1 & 1 & 1 & 0 & 0 & 0 & 0 & | & 0 \\ 
    0 & 1 & 1 & 1 & 0 & 1 & 0 & | & 0 \\ 
    1 & 0 & 1 & 1 & 1 & 0 & 1 & | & 1 
    \end{bmatrix}. 
\end{equation} 
We then continue to execute \textbf{\textit{Step~2}} and \textbf{\textit{Step~3}}, and find that when $(n,k)=(3,3)$, we can complete the elimination and satisfy \eqref{eq:consistency_check}. The augmented matrix at this  time is as follows:
\begin{equation}\label{eq:matrix_1}
    [H_{\pi}^{(3)} \mid \underline{s}^{(3)}] = 
    \begin{bmatrix} 
    1 & 0 & 0 & 1 & 0 & 1 & 1 & | & 0 \\ 
    0 & 1 & 0 & 1 & 1 & 0 & 1 & | & 1 \\ 
    0 & 0 & 1 & 0 & 1 & 1 & 1 & | & 1
    \end{bmatrix}. 
\end{equation}
From \eqref{eq:matrix_1}, since the target syndrome is $\underline{s}^{(3)} = (0,1,1)^{\mathsf{T}}$, the corresponding solution supported on the first three columns is $\mathcal{E}_{3} = \{(0,1,1,0,0,0,0)^{\mathsf{T}}\}$.

\begin{table}[htbp]
    \centering
    \renewcommand{\arraystretch}{1.1}
    \begin{tabular}{|c|c|c|c|}
    \hline
    Priority $t$ & EP $\tilde{\underline{e}}(t)^{\mathsf{T}}$ & RMRE & Status \\
    \hline
    1 & $(1,0,0,0,0,0,0)$ & 1 & Filtered out \\
    2 & $(0,1,0,0,0,0,0)$ & 2 & Filtered out \\
    3 & $(0,0,1,0,0,0,0)$ & 3 & Filtered out \\
    4 & $(1,1,0,0,0,0,0)$ & 2 & Filtered out \\
    5 & $(0,0,0,1,0,0,0)$ & 4 & Verified \\
    6 & $(1,0,1,0,0,0,0)$ & 3 & Filtered out \\
    7 & $(0,0,0,0,1,0,0)$ & 5 & Verified (output) \\
    8 & $(1,0,0,1,0,0,0)$ & 4 & Pending \\
    9 & $(0,1,1,0,0,0,0)$ & 3 & Determined by GE \\
    \hline
    \end{tabular}
    \vspace{1mm}
    \caption{Screening process of EPs for the $(7,4)$ Hamming code with updated status descriptions.}
    \label{tab:example_screening}
\end{table}

Then we proceed to \textbf{\textit{Step~4}}. The screening process is summarized in Table \ref{tab:example_screening}, where the first three columns have been pre-stored locally.  The EP determined by GE is $\tilde{\underline{e}}(9) = (0,1,1,0,0,0,0)^{\mathsf{T}}$. Then, we sequentially check the candidate EPs: the EPs with RMRE $\leq$ 3 are filtered out, so only $\tilde{\underline{e}}(5)$ and $\tilde{\underline{e}}(7)$ need to be verified. The $\tilde{\underline{e}}(7)$ can be determined as a valid EP, so we output it and confirm that the final decoded codeword is $\theta(\underline{y}) \oplus \mathcal{P}\cdot \tilde{\underline{e}}(7)$. In this example, the number of guesses is reduced from $7$ to $2$ (only $\tilde{\underline{e}}(5)$ and $\tilde{\underline{e}}(7)$).

\subsection{Complexity Analysis}\label{subsec:complexity}

The proposed algorithm can be divided into two parts: partial GE and filtered ORBGRAND. This subsection analyzes its computational complexity and presents an optimized design that further reduces the cost of the GE process.

We first consider the GE complexity in \textbf{Algorithm~\ref{alg:filter_GRAND}}.  
Each elimination round in \eqref{eq:elimination} requires approximately $(N+1)(N-K)$ operations.  
Since only a partial elimination is performed, when $k$ rounds of elimination are completed, we record the end round $m=k$, and the maximum complexity is
\begin{equation}
    \mathcal{O}_{\text{Partial GE}} = m\,(N+1)(N-K).
\end{equation}
However, only the first $m$ bits are relevant for constructing valid EPs.  
Once a feasible RMRE has been found, elimination on later columns becomes redundant.  
Therefore, we simplify \textbf{\textit{Step~2}} in the previous subsection \textbf{Algorithm~\ref{alg:filter_GRAND}} as follows.

\begin{enumerate}
    \item \textbf{First pivot:}  $(n,k) = (1,1)$, find a nonzero pivot $h_{p1}^{(0)}$ in $\underline{h}_1^{(0)}$. After swapping it to the first row, record the row transformation matrix $(I + Q^{(1)})$, where $Q^{(1)}_{i1}=h_{i1}^{(0)}$ and other positions are set to zero. 
    The operation applies only to the first column and syndrome vector:
    \begin{equation}\label{eq:first_pivot}
        h_{i1}^{(1)}=0,\quad s_i^{(1)}=s_i^{(0)}\oplus h_{i1}^{(0)}s_1^{(0)},\quad \forall\ i\neq1.
    \end{equation}
    Then set $k\gets k+1$ and proceed to \textbf{\textit{Step~3}}.
    
    \item \textbf{Subsequent pivots:} Before processing the $n$-th column, apply the accumulated row transformations:
    \begin{equation}\label{eq:accumulated_transform}
        \underline{h}_{n}^{(k-1)} \leftarrow 
        \Biggl(\prod_{j=k-1}^{1}(I + Q^{(j)})\Biggr)\underline{h}_{n}^{(0)}.
    \end{equation}
    GE is then performed on the updated column $\underline{h}_{n}^{(k-1)}$: If a pivot is found, perform elimination on $\underline{h}_{n}^{(k-1)}$ and $\underline{s}^{(k-1)}$ as in \eqref{eq:first_pivot}, store the new transformation matrix $(I + Q^{(k)}),\ k\gets k + 1$ and proceed to \textbf{\textit{Step~3}}; otherwise, move to $(n+1,k)$.
\end{enumerate}

\begin{example}\label{ex:2}
    We demonstrate the original and reduced partial GE procedures on the matrix \eqref{eq:matrix_0} of \textit{Example \ref{ex:1}}, as shown in Figure~\ref{fig:GE}.

    In the full GE method (Figure~\ref{fig:GE}(a1) - (a4)), all columns are updated in each round.
    In contrast, the proposed scheme (Figure~\ref{fig:GE}(b1) - (b4)) only updates the active column and $\underline{s}$.
    Before processing the next column, the accumulated transformation is applied once,  
    avoiding redundant operations while preserving correct pivot formation.
\end{example}

\smallskip
We now estimate the complexity of this simplified process. In the first round, the computational cost in \eqref{eq:first_pivot} is at most $2(N-K)$.
For the $k$-th pivot, the accumulated transformation in the \eqref{eq:accumulated_transform} and elimination together require
\begin{equation*}
    (k-1)(N-K) + 2(N-K) = (k+1)(N-K)\ \text{operations}.
\end{equation*}
If $m$ pivots are found in total, the overall complexity becomes
\begin{equation*}
    \mathcal{O}_{\text{Reduced GE}}
    = \sum_{k=1}^{m}\left[(k+1)(N-K)\right]
    = \frac{m(m+3)}{2}(N-K).
\end{equation*}

Finally, we summarize the overall decoding complexity.  
Let $T_1$ denote the number of guesses required by the original ORBGRAND, and let the verification cost per trial be $cN$, where $c$ represents the average number of parity-check equations used during verification.  
For the proposed elimination-aided scheme, the number of required guesses is denoted by $T_2$, where $T_2 \leq T_1$ due to the filtering mechanism.  
Considering the additional cost of the partial GE stage, the total decoding complexity can be expressed as
\begin{align}
     \mathcal{O}_{\text{ORBGRAND}} &= T_1 cN, \label{eq:complexity_1}\\
     \mathcal{O}_{\text{With partial GE}} &= \mathbb{E}\left[M\right](N+1)(N-K) + T_2 cN, \label{eq:complexity_2}\\
     \mathcal{O}_{\text{With reduced GE}} &= \mathbb{E}\left[\frac{M(M+3)}{2}\right](N-K) + T_2 cN. \label{eq:complexity_3}
\end{align}

\subsection{Discussion}\label{subsec:discuss}

From \eqref{eq:complexity_1}–\eqref{eq:complexity_3}, it can be seen that the effectiveness of the proposed algorithm depends on the tradeoff between the additional cost introduced by elimination and the complexity reduction achieved by filtering out some EPs.  

To further understand this tradeoff, we analyze the statistical property of the $M$ through $\text{RMRE}(\theta(\underline{Y}) \oplus \underline{W})$.  
Since for each received vector, we have
\begin{equation}
    m  \leq \min\{\text{RMRE}(\tilde{\underline{e}}) \mid H_{\pi}\cdot \tilde{\underline{e}} = \underline{s}\},
\end{equation}
and the $\theta(\underline{y}) \oplus \underline{w}$ is the target EP as in \eqref{eq:rmre_2}, then we have $m \leq \text{RMRE}(\theta(\underline{y}) \oplus \underline{w})$, and for any monotonically increasing function $g(\cdot)$:
\[
\mathbb{E}[g(M)] \leq \mathbb{E}[g(\text{RMRE}(\theta(\underline{Y}) \oplus \underline{W}))].
\]
The distribution of \(\text{RMRE}(\theta(\underline{Y}) \oplus \underline{W})\) can then be characterized using the theory of order statistics, as derived in \cite{song2024fundamentals}.
Specifically, let
    $P_i = \Pr\big(\text{RMRE}(\theta(\underline{Y}) \oplus \underline{W}) = i\big)$.
Then under the symmetry channel, we have $P_0 = \big(1 - F(0)\big)^N$, and
\begin{align}\label{eq:Pi_symmetric} 
    P_i = \frac{N!}{(i-1)!(N-i)!}  \int_0^\infty &\big[F(t)-F(-t)\big]^{i-1} \notag \\ 
    & \cdot  f(-t)\,\big[1 - F(t)\big]^{N-i} \, \mathrm{d}t.
\end{align}
Where $F(\cdot)$ and $f(\cdot)$ are the cumulative distribution function and probability density function of $L$ when $X= + 1$.


Thus, when channel statistics cause $\text{RMRE}(\theta(\underline{Y}) \oplus \underline{W})$ to be concentrated at a small rank, the expected value $\mathbb{E}[M]$ is much smaller than the code length $N$. Thus, the additional cost of partial GE remains small compared to the parity check operation in our setup.

\section{Performance Evaluation}\label{sec:performance}

This section presents simulation results that verify the effectiveness of the proposed algorithm using the BCH(127,113) code.  
The maximum number of guesses is set to $T = 5\times10^4$, and the primary decoding methods evaluated are ORBGRAND \cite{duffy2022ordered} and its recent improved variant RS-ORBGRAND \cite{wan2024approaching}. 

\begin{figure}[htbp]
    \centering
    \includegraphics[width=0.96\linewidth]{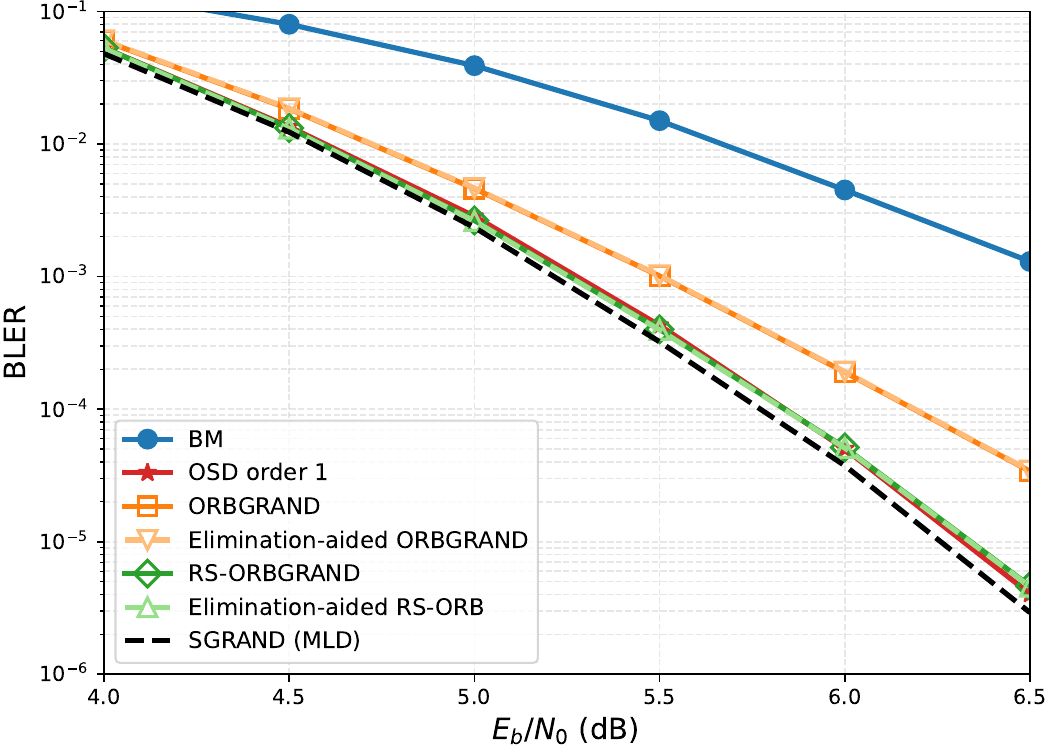}
    \caption{Comparison of BLER curves under different $E_b/N_0$s.}
    \label{fig:inv_count1}
\end{figure}

Figure~\ref{fig:inv_count1} compares the decoding performance of various decoders on BCH (127, 113). These include Berlekamp-Massey (BM) decoding \cite{berlekamp2015algebraic}, OSD, and multiple GRAND variants. Among them, soft GRAND (SGRAND) \cite{solomon2020soft} realizes maximum likelihood decoding (MLD)\footnote{MLD is achieved when the decoder is permitted to enumerate all EPs without imposing any cap on the maximum number of guesses.}. We apply our acceleration algorithm to both ORBGRAND and RS-ORBGRAND, respectively. We show that the elimination-aided ORBGRAND achieves the same block error rate (BLER) compared to the original ORBGRAND algorithm. Furthermore, despite relying only on the ranking value of soft information, the RS-ORBGRAND still performs very close to MLD and achieves nearly the same performance as the OSD of order 1.

\begin{figure}[htbp]
    \centering
    \includegraphics[width=0.96\linewidth]{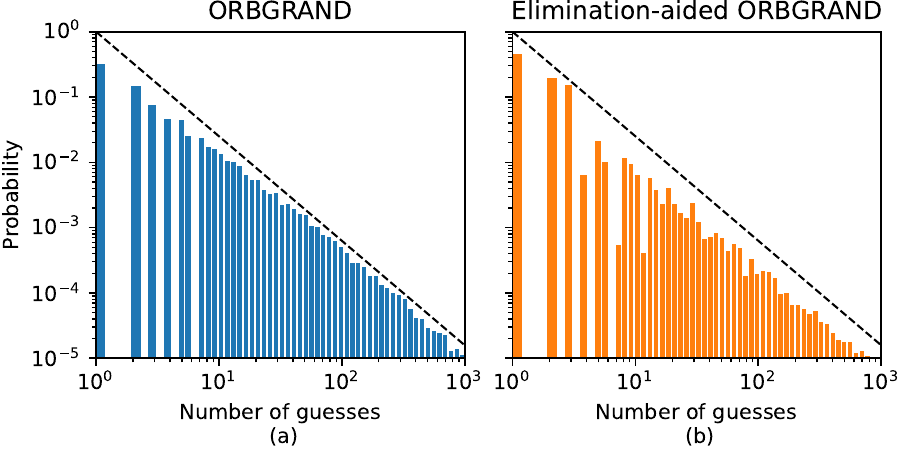}
    \caption{Histogram of number of guesses for BCH(127, 113), $E_b/N_0$ = 5 dB.}
    \label{fig:query}
\end{figure}

Table~\ref{tab:simulation_results} summarizes the average number of guesses obtained after applying our accelerated algorithm, showing that the proposed method consistently reduces the number of guesses by approximately 40\%–50\% for various $E_b/N_0$ values. Figure~\ref{fig:query} further presents the distribution of number of guesses. It can be observed that the proposed approach effectively concentrates the guesses around smaller values, thereby significantly reducing the average number of guesses and mitigating the issue of high tail latency.

\begin{table}[htbp]
    \centering
    \renewcommand{\arraystretch}{1.16}
    \begin{tabular}{|l|
    >{\centering\arraybackslash}p{1.1cm}|
    >{\centering\arraybackslash}p{1.1cm}|
    >{\centering\arraybackslash}p{1.1cm}|
    }
        \hline
        \multirow{2}{*}{\textbf{Decoding Method}} 
        & \multicolumn{3}{c|}{\textbf{Average Guesses at Each $E_b/N_0$}} \\ \cline{2-4}
        & \textbf{4 dB} & \textbf{5 dB} & \textbf{6 dB} \\ 
        \hline
        ORBGRAND & 1.04e+3 & 9.67e+1 & 7.30e+0 \\
        Elimination-aided ORBGRAND & 5.92e+2 & 4.49e+1 & 3.27e+0 \\
        \rowcolor{gray!12}\textbf{Reduction in Guesses (\% $\downarrow$)} & \textbf{43.1} & \textbf{53.5} & \textbf{55.2} \\
        \hline
        RS-ORB & 9.52e+2 & 7.02e+1 & 4.52e+0 \\
        Elimination-aided RS-ORB & 6.09e+2 & 4.16e+1 & 2.40e+0 \\
        \rowcolor{gray!12}\textbf{Reduction in Guesses (\% $\downarrow$)} & \textbf{36.0} & \textbf{40.7} & \textbf{46.9} \\
        \hline
    \end{tabular}
    \vspace{0.5mm}
    \caption{Comparison of average number of guesses for different decoding methods}
    \label{tab:simulation_results}
\end{table}

\begin{figure}[htbp]
    \centering
    \includegraphics[width=0.96\linewidth]{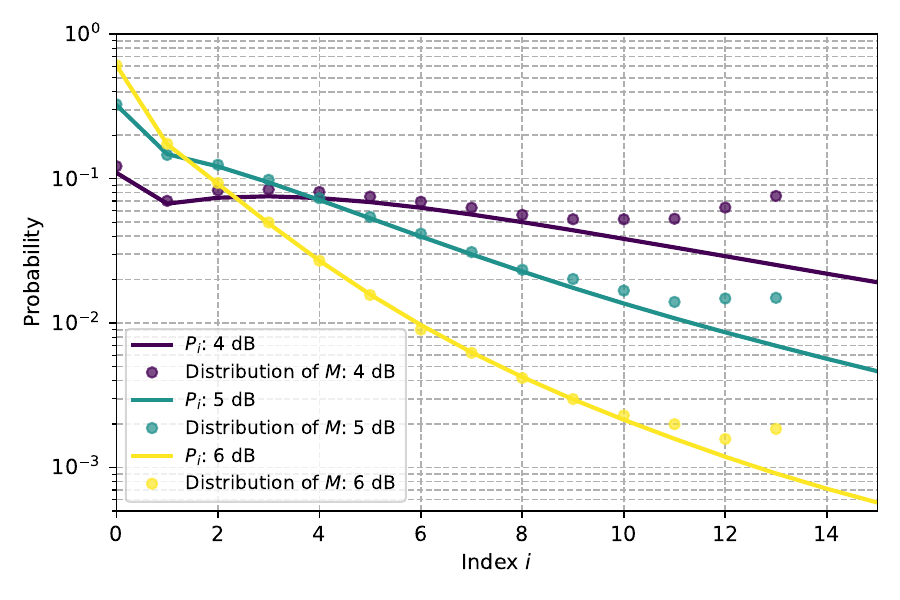}
    \caption{Distribution of $\text{RMRE}(\theta(\underline{Y}) \oplus \underline{W})$ and $M$.}
    \label{fig:rmre_dist}
\end{figure}

Regarding the complexity analysis of the partial GE, we first validate the results presented in Section~\ref{subsec:discuss}. Figure~\ref{fig:rmre_dist} illustrates the distributions of the $\text{RMRE}(\theta(\underline{Y}) \oplus \underline{W})$ and $M$, respectively. The RMRE values are derived from \eqref{eq:Pi_symmetric}, whereas $M$ is estimated via Monte Carlo sampling over repeated decoding trials. The results indicate that most RMRE values are relatively small, suggesting that for the considered setup, successful decoding typically requires only a few bit flips.

\begin{figure}[htbp]
    \centering
    \includegraphics[width=0.99\linewidth]{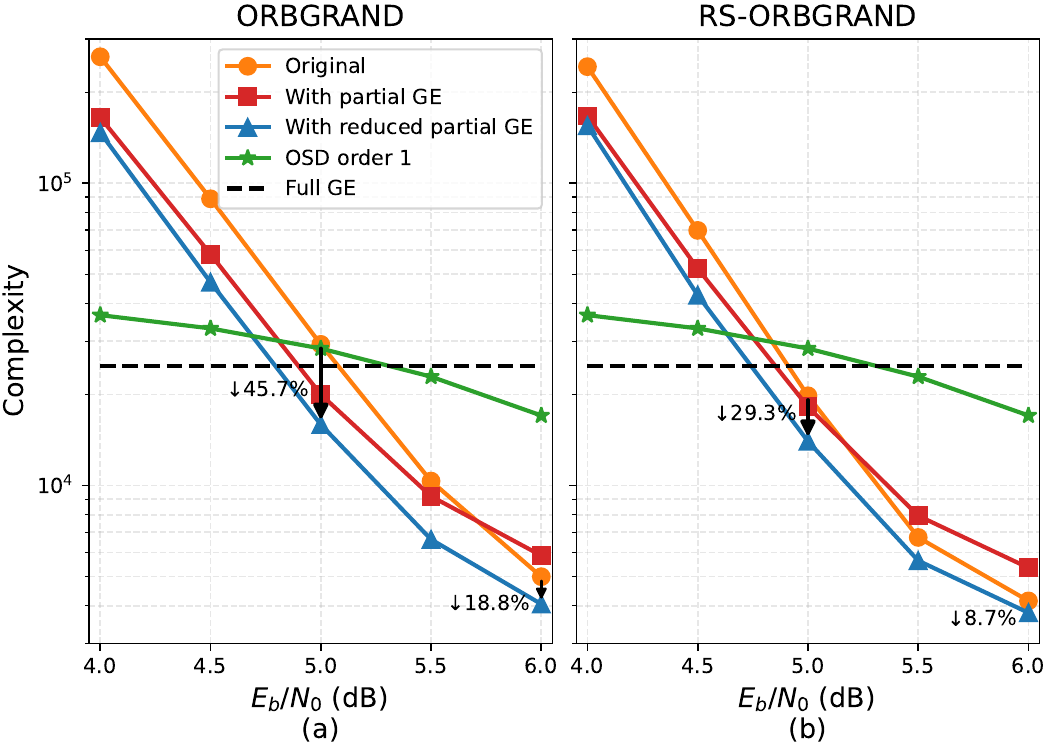}
    \caption{Comparison of average computational complexity under different $E_b/N_0$ values (1 floating-point operation = 8 XORs).}
    \label{fig:inv_count2}
\end{figure}

Figures~\ref{fig:inv_count2}(a) and (b) compare the decoding complexity of ORBGRAND and RS-ORBGRAND under different implementations. The vertical axis reports the average number of XOR operations, where one floating-point operation is counted as eight XORs. Although partial GE can reduce complexity at low $E_b/N_0$, its benefit diminishes—and may even vanish—at high $E_b/N_0$, since each elimination step still operates on full rows. In contrast, the reduced partial GE method further lowers the elimination complexity, leading to notable improvements in overall performance.Meanwhile, full GE remains expensive, which limits the complexity reduction of OSD as $E_b/N_0$ increases. Consequently, ORBGRAND preserves a clear complexity advantage in the high-$E_b/N_0$ regime.

At $E_b/N_0=5$ dB in Fig.~\ref{fig:inv_count2}, a $53.5\%$ and $40.7\%$ reduction in the number of guesses (see Table~\ref{tab:simulation_results}) translates into a $45.7\%$ and $29.3\%$ reduction in overall decoding complexity, respectively. Even at $6$ dB, where sorting becomes a larger fraction of the total cost due to very few guesses, we still observe speedups of $18.1\%$ and $8.7\%$, confirming that removing redundant EPs by reduced partial GE yields measurable end-to-end latency reductions.

\section{Conclusion}\label{sec:conclusion}

We proposed an elimination-aided ORBGRAND scheme that exploits the RMRE to accelerate decoding. By integrating partial GE, multiple EPs with the same RMRE are verified jointly, thereby filtering out redundant EP checks and lowering complexity of GE from $\mathcal{O}(N(N-K)^2)$ to $\mathcal{O}((N-K)M^2)$. Simulations show that the method significantly speeds up decoding while maintaining the same error-correction capability. The approach is hardware-friendly and can be seamlessly integrated with OSD for rare, hard-to-decode cases, offering a scalable framework for GRAND-based architectures.

\bibliographystyle{IEEEtran}
\bibliography{Ref}

@article{yue2023efficient,
  title={Efficient decoders for short block length codes in {6G} {URLLC}},
  author={Yue, Chentao and Miloslavskaya, Vera and Shirvanimoghaddam, Mahyar and Vucetic, Branka and Li, Yonghui},
  journal={IEEE Communications Magazine},
  volume={61},
  number={4},
  pages={84--90},
  year={2023},
  publisher={IEEE}
}

@book{song2024fundamentals,
  title={Fundamentals of Order and Rank Statistics},
  author={Song, Iickho and Park, So Ryoung and Zhang, Wenyi and Lee, Seungwon},
  year={2024},
  publisher={Springer}
}

@book{berlekamp2015algebraic,
  title={Algebraic Coding Theory},
  author={Berlekamp, Elwyn R},
  year={2015},
  publisher={World Scientific}
}

@article{duffy2019capacity,
  title={Capacity-achieving guessing random additive noise decoding},
  author={Duffy, Ken R and Li, Jiange and M{\'e}dard, Muriel},
  journal={IEEE Transactions on Information Theory},
  volume={65},
  number={7},
  pages={4023--4040},
  year={2019},
  publisher={IEEE}
}

@inproceedings{solomon2020soft,
  title={Soft maximum likelihood decoding using {GRAND}},
  author={Solomon, Amit and Duffy, Ken R and M{\'e}dard, Muriel},
  booktitle={IEEE International Conference on Communications (ICC)},
  pages={1--6},
  year={2020}
}

@article{duffy2022ordered,
  title={Ordered reliability bits guessing random additive noise decoding},
  author={Duffy, Ken R and An, Wei and M{\'e}dard, Muriel},
  journal={IEEE Transactions on Signal Processing},
  volume={70},
  pages={4528--4542},
  year={2022},
  publisher={IEEE}
}

@article{liu2022orbgrand,
  title={{ORBGRAND} is almost capacity-achieving},
  author={Liu, Mengxiao and Wei, Yuejun and Chen, Zhenyuan and Zhang, Wenyi},
  journal={IEEE Transactions on Information Theory},
  volume={69},
  number={5},
  pages={2830--2840},
  year={2022},
  publisher={IEEE}
}

@article{abbas2022high,
  title={High-throughput and energy-efficient {VLSI} architecture for ordered reliability bits {GRAND}},
  author={Abbas, Syed Mohsin and Tonnellier, Thibaud and Ercan, Furkan and Jalaleddine, Marwan and Gross, Warren J},
  journal={IEEE Transactions on Very Large Scale Integration (VLSI) Systems},
  volume={30},
  number={6},
  pages={681--693},
  year={2022},
  publisher={IEEE}
}

@article{ji2024efficient,
  title={Efficient {ORBGRAND} Implementation With Parallel Noise Sequence Generation},
  author={Ji, Chao and You, Xiaohu and Zhang, Chuan and Studer, Christoph},
  journal={IEEE Transactions on Very Large Scale Integration (VLSI) Systems},
  year={2024},
  publisher={IEEE}
}

@article{fossorier1995soft,
  title={Soft-decision decoding of linear block codes based on ordered statistics},
  author={Fossorier, Marc PC and Lin, Shu},
  journal={IEEE Transactions on information Theory},
  volume={41},
  number={5},
  pages={1379--1396},
  year={1995},
  publisher={IEEE}
}

@INPROCEEDINGS{wan2024approaching,
  author={Wan, Li and Zhang, Wenyi},
  booktitle={IEEE International Symposium on Information Theory (ISIT)}, 
  title={Approaching Maximum Likelihood Decoding Performance via Reshuffling {ORBGRAND}}, 
  year={2024},
  volume={},
  number={},
  pages={31-36},
  doi={10.1109/ISIT57864.2024.10619402}
}

@inproceedings{rowshan2022constrained,
  title={Constrained error pattern generation for {GRAND}},
  author={Rowshan, Mohammad and Yuan, Jinhong},
  booktitle={IEEE International Symposium on Information Theory (ISIT)},
  pages={1767--1772},
  year={2022}
}

@inproceedings{hadavian2023ordered,
  title={Ordered reliability direct error pattern testing ({ORDEPT}) algorithm},
  author={Hadavian, Reza and Truhachev, Dmitri and El-Sankary, Kamal and Ebrahimzad, Hamid and Najafi, Hossein},
  booktitle={IEEE Global Communications Conference},
  pages={6983--6988},
  year={2023}
}

@inproceedings{wang2024partially,
  title={Partially constrained {GRAND} of linear block codes},
  author={Wang, Yixin and Liang, Jifan and Ma, Xiao},
  booktitle={9th International Conference on Computer and Communication Systems (ICCCS)},
  pages={617--622},
  year={2024}
}

@article{wan2025parallelism,
  title={A Parallelization Strategy for {GRAND} with Optimality Guarantee by Exploiting Error Pattern Tree Representation},
  author={Wan, Li and Yin, Huarui and Zhang, Wenyi},
  journal={arXiv preprint arXiv:2510.01813},
  year={2026}
}






\end{document}